\documentclass[fleqn,twoside]{article}
\usepackage{espcrc2}

\usepackage{graphicx}
\usepackage[figuresright]{rotating}

\newcommand{\AmS}{{\protect\the\textfont2
  A\kern-.1667em\lower.5ex\hbox{M}\kern-.125emS}}

\title{Chiral Limit of Staggered Fermions at Strong Couplings: \\
	A Loop Representation}

\author{Shailesh Chandrasekharan
	\address{Department of Physics, Duke University, Durham, NC 27708-0305}
        \thanks{This work was done in collaboration with David Adams and
		was partially supported by the US Department of Energy
		grant DE-FG02-96ER40945.}}
       
\begin{document}

\begin{abstract}
The partition function of two dimensional massless staggered 
fermions interacting with $U(N)$ gauge fields is rewritten
in terms of loop variables in the strong coupling limit.
We use this representation of the theory to devise a non-local
Metropolis algorithm to calculate the chiral susceptibility. 
For small lattices our algorithm reproduces exact results 
quite accurately. Applying this algorithm to large volumes
yields rather surprising results. In particular we find 
$m_\pi \neq 0$ for all $N$ and it increases with $N$. 
Since the talk was presented we have found reasons to believe 
that our algorithm breaks down for large volumes questioning
the validity of our results.
\vspace{1pc}
\end{abstract}
\maketitle

\section{INTRODUCTION}

Understanding the chiral limit of lattice gauge theories 
with fermions still presents a formidable challenge for 
numerical work. Conventional algorithms like the hybrid 
Monte-Carlo or multi-boson techniques encounter 
difficulties in this limit. In this article we discuss
how a new loop representation helps one to solve the chiral 
limit of strongly coupled $U(N)$ gauge theories interacting 
with staggered fermions in two dimensions. 

The chiral limit of strongly coupled gauge theories involving
staggered fermions in two dimensions contain some unresolved 
questions. It is well known that staggered fermions preserve a 
$U(1)$ chiral symmetry at finite lattice spacings. Using 
large $N$ (color) \cite{Kaw81} and large $d$ (dimension) \cite{Klu81} 
approximations, it is possible to show that this chiral symmetry 
breaks spontaneously at strong couplings. There are also some 
rigorous results in four dimensions \cite{Sal91}. As a consequence, 
at least in $d\geq 3$ the model is expected to contain a single 
Goldstone boson (pion) which is exactly massless in the chiral 
limit. What happens in two dimensions? Numerical simulations 
suggest the existence of a massless pion \cite{Gut99}. This is 
somewhat surprising since in $d=2$ the Mermin-Wagner theorem 
forbids the breaking of a continuous symmetry and the pion cannot 
be a Goldstone boson. On the other hand in two dimensions we know 
that a $U(1)$ symmetry is special. Kosterlitz and Thouless have
shown that a theory with this symmetry can contain both a massive 
phase and a massless phase separated by a transition. 
So an interesting question is whether the strongly coupled $U(N)$ 
lattice gauge theory with staggered fermions contains a massless 
pion in the chiral limit or not and if it does is the long distance
physics in the same universality of the 2d X-Y model.

\section{LOOP REPRESENTATION}

About two decades ago, it was shown that a $U(N)$ gauge theory 
involving a single flavor of staggered fermions in the strong coupling 
limit in any dimension is equivalent to a statistical mechanics of 
monomers and dimers \cite{Wol84}. The partition function is given by
\begin{equation}
Z \;=\; \sum_{[n,b]} \;\;\;\;\;
\prod_{x,\mu}\;\frac{(N-b_{x,\mu})!}{b_{x,\mu}! N!}\;\;\; 
\prod_x \frac{N!}{n_x!}\;m^{n_x}
\label{bpf}
\end{equation}
where $x$ represents a lattice point in the $d$ dimensional
lattice, $\mu = 1,2,...d$ the direction, $\hat{\mu}$ the 
unit vector in the corresponding direction, $n_x=0,1,...,N$ is 
the number of monomers located on the site $x$ and 
$b_{x,\mu}=0,1,2,...,N$ is the number of dimers located on the 
bond connecting the site $x$ and the nearest neighbor site 
$x+\hat{\mu}$ in the $\mu$ direction. We will assume 
$b_{x,\mu}\;\equiv\; b_{x+\hat{\mu}, -\mu}$ for simplicity. 
The configurations $[n,b]$ that contribute to the partition 
function are constrained by the relation
\begin{equation}
n_x + \sum_\mu \; b_{x,\mu} + b_{x,-\mu} \;=\; N
\end{equation}
on each site $x$. 

\begin{figure}[htb]
\vskip-0.2in
\begin{center}
\includegraphics[width=0.35\textwidth]{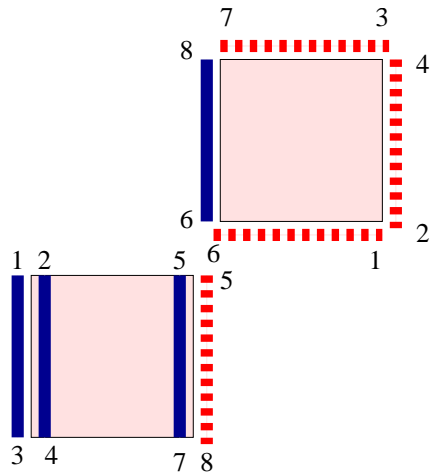}
\caption{ \label{2x2loopc} \em An $N=2$ loop configuration 
on a $2\times 2$ periodic lattice. The two plaquettes are 
shown separately for convenience. The numbers at the end of 
the dimers indicate connections. There are two loops in this
configuration.}
\end{center}
\vskip-0.2in
\end{figure}

In the chiral limit $n_x$ is forced to be zero and it is difficult 
to design a local algorithm for the partition function that obeys 
the constraint. Perhaps for this reason the chiral limit remained 
unexplored. In order to avoid this problem we rewrite the partition 
function given in (\ref{bpf}) in terms of new ``loop'' variables.
For the case of $N=1$ this was recently explained in \cite{Cha01}.
We first shade plaquettes of the two dimensional lattice as in a 
chess board. We then introduce new (dashed) dimers 
$d_{x,\mu}=0,1,2...N$ on each bond such that each corner $x$ of 
any shaded plaquette satisfies
\begin{equation}
b_{x,\mu} + d_{x,\mu} + b_{x,\nu} + d_{x,\nu} = N
\label{dbc}
\end{equation}
where $x+\hat{\mu}$ and $x + \hat{\nu}$ are the two nearest neighbor 
sites of $x$ also belonging to the same shaded plaquette. In
(\ref{dbc}) $\mu$ and $\nu$ can be negative if the direction of the 
nearest neighbor site is backward. Of course there are many ways to 
introduce the $d_{x,\mu}$ bonds that satisfy (\ref{dbc}). It is 
easy to check that for a shaded plaquette with bonds $b1,b2,b3,b4$ 
in a cyclic order, the total number of ways one can introduce the 
dashed bonds obeying the constraint (\ref{dbc}) is 
$(N+1 - \mbox{max}(b1,b3) - \mbox{max}(b2,b4))$. Thus in order to 
keep the partition function the same we re-weight each extended 
configuration with $1/(N+1 - \mbox{max}(b1,b3) - \mbox{max}(b2,b4))$.
In this construction it is easy to see that each site touches
$N-n_x$ solid (original) dimers and $N+n_x$ dashed dimers. When
$n_x=0$ one can at random connect every solid dimer to a unique 
dashed dimers in $N!$ ways. In the chiral limit this results in 
loops made up of an alternate sequence of solid and dashed dimers. 
In figure \ref{2x2loopc} an $N=2$ configuration of solid dimers on a
$2\times 2$ lattice  is extended to a loop configuration using the
above rules.
An update of the system involves flipping of a loop which means
converting all solid dimers into dashed dimers within the loop
and vice versa. Typically, a point is selected at random and the 
loop connected to it is flipped using a Metropolis decision. Since 
there are $(b+d)!/(d! b!)$ number of equivalent dimer configurations 
with $b$ solid and $d$ dashed dimers on a given bond, it is necessary 
to take this factor into account in the Metropolis decision to 
ensure detailed balance. 

\section{CHIRAL SUSCEPTIBILITY}

To address the question of the existence of a massless pion in the 
chiral limit one can measure the chiral susceptibility $\chi$ as a 
function of the lattice size $L$. Since it is the first derivative of the 
chiral condensate with respect to the fermion mass it should
diverge as a power of $L$ if there is a massless pion in the
theory. For example if the chiral symmetry is spontaneously broken 
we must have $\chi \propto L^d$ where $d$ is the dimensions
of the lattice. In the case of two dimensions where we 
expect chiral symmetry to remain unbroken, a massless
pion leads to a divergence of the form $\chi \propto L^(2-\eta)$
where the KT prediction is that $0 \leq \eta \leq 0.25$. Another
example is the continuum two flavor Schwinger model where 
$\eta = 1$.

To measure the chiral susceptibility, we introduce two monomers 
into each loop configuration in all possible ways. Since
the loops are made up of an alternate sequence of solid
and dashed bonds, introducing a monomer introduces a defect
in the pattern. However, a single defect is impossible in
a loop. Thus, in a two monomer configuration both monomers must 
be on the same loop. It is easy to see that a loop with two
monomers can be generated by partially flipping a loop. Using this 
observation we find the chiral susceptibility by summing the 
weights of all the new configurations generated during the loop 
update. We have checked that our measurement of the susceptibility 
produces accurate answers on small lattices. Table 1, compares our 
results to exact calculations.

\begin{table}[ht]
\vskip-0.2in
\begin{center}
\caption{ {\label{exact} Susceptibility: algorithm vs. exact results.}}
\vskip0.1in
\begin{tabular}{|c|c|l|c|}
\hline
N & Lattice Size & Exact & Algorithm \\
\hline
1 & $8 \times 8$ &    5.27221... & 5.2722(2) \\
2 & $4 \times 4$ &    6.40205... & 6.402(2) \\
3 & $4 \times 4$ &    14.1595... &14.159(8) \\
5 & $2 \times 2$ &    9.57386... & 9.574(7) \\
30 & $2 \times 2$ &   338.534... & 338.2(8) \\
\hline
\end{tabular}
\end{center}
\vskip-0.3in
\end{table}

\begin{figure}[htb]
\begin{center}
\includegraphics[width=0.45\textwidth]{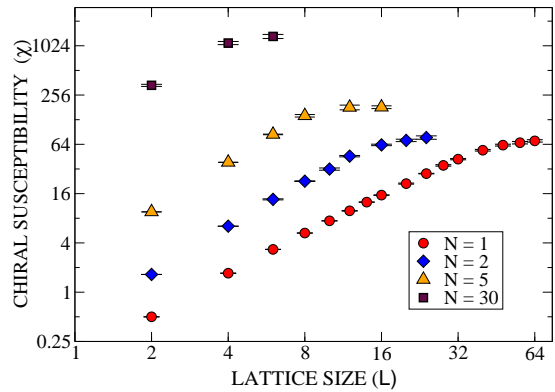}
\caption{ \label{susN} \em Chiral Susceptibility as a function
of the lattice size $L$ for various values of $N$.}
\end{center}
\vskip-0.2in
\end{figure}

When we apply this algorithm to larger lattices we find that
small clusters often get updated and occasionally even loops of 
large sizes get flipped. Our results for the chiral 
susceptibility for various lattice sizes is plotted in 
figure \ref{susN}.
We can draw three conclusions from the plot. First, as a function 
of $L$ there is no evidence of a divergence for any $N$.
Thus the pion is always massive. Second, since the value of $L$ 
where the susceptibility begins to level gives a rough indication 
of the inverse of the pion mass, the pion mass increases with $N$. 
Finally, the value of the chiral susceptibility diverges with 
$N$, signaling the formation of a condensate. However, our results 
suggest that this is not driven by long distance fluctuations. 

Since the talk was given we have continued to study our algorithm
more carefully. It now appears to us that there are configurations 
which we sample exponentially rarely as the volume increases but 
whose contribution is still finite to the final result. This problem
begins to show up mildly around $L=32$ for $N=1$ through very large 
fluctuations.
Of course we had collected very very large statistics to be able 
to control our error bars. However, we think the error analysis is
not reliable since we may have uncontrolled auto-correlation times.
For this reason we believe the error bars in figure \ref{susN} and 
in the graphs presented in \cite{Cha01} are incorrect. This also
invalidates the conclusions made above and in \cite{Cha01}.
Thus, although our representation is a useful first step and perhaps 
gives a qualitative picture for small volumes more work is needed 
before we can answer questions we raised in the first section.

\end{document}